# Allelopathy of *Rumex* spp.: A review

**Aram Akram Mohammed**
*Department of Horticulture, College of Agricultural Engineering Sciences, University of Sulaimani, Sulaymaniyah, Kurdistan Region, Iraq.*

**Abstract:**
The genus of *Rumex* from the Polygonaceae family is widespread in the world, particularly in the northern hemisphere, and includes about 250 species. The species of this genus are used for medicinal purposes and their allelopathic impacts. Regarding allelopathy, many allelochemicals have been detected in different *Rumex* species. Therefore, plant extracts, leachates, and plant residues of different species of *Rumex* have been studied with seed germination and plant growth in the recipient plants. Also, various species of *Rumex* were tested for their allelopathic capacities to control weeds, insects, and plant pathogens. Besides, it was revealed that the allelopathic impact of *Rumex* spp. was variable depending on extract concentration, the plant part of the *Rumex* spp., and the species of the recipient plant. In the present review, the results of the studies are exhibited that aimed at the allelopathic effect on different aspects of the plant crops, weeds, insects, and plant pathogens.

## 1- Introduction

Allelopathy is a dual term that, at the same time, covers both inhibitory and stimulatory consequences of certain plant activities on reproduction and growth of other plants or nearby organisms and animals (Grodzinskii, 2016). This term was first coined by Molisch (1937) in his book "The influence of one plant on another: Allelopathy". The allelopathic effect of plants, on one hand, has been investigated *in vivo* through the study of chemicals released by roots, the decomposition of leaf litter, or volatilization that may have an allelopathic impact on the neighboring plants; on the other hand, via the study of plant extracts or isolated chemicals and the application of them *in vitro* or *ex vitro* to seeds, organs, or intact plants (Singh et al., 2009; Zhu et al., 2011). Plants exert allelopathic phenomena via specific chemicals, which are known as allelochemicals; in most cases, they are secondary metabolites (Willis, 2007). In agriculture, allelopathic results of plants have been exploited for some purposes. Application of allelopathic may be helpful for weed management, pest and disease control, improving crop productivity, and maintaining soil fertility (Cheng and Cheng, 2015). Allelopathic principle incorporation in agricultural systems encourages sustainable practices because it reduces synthetic chemical dependence and protects ecosystem and human health (Khanh et al., 2005).

---

**Corresponding author*:** E-mail address: aram.hamarashed@univsul.edu.iq

---

Many plants have been tested for their allelopathic effects on other plants, including *Rumex* spp. (Mardani et al., 2016).

The genus of *Rumex* has common names of docks and sorrels, which belong to the Polygonaceae family. This genus is widespread in the world, particularly in the northern hemisphere, and includes about 250 species (Rao et al., 2011). The members of the genus are mostly perennial, whereas there are several species that exhibit biennial or annual life cycles (Anjen et al., 2003). They have thrived and grown in diverse habitats, such as disturbed areas, roadsides, fields, meadows, and wetland margins (Van Assche et al., 2002). However, several species are notorious weeds, but many of them are used for medicinal purposes and their allelopathic impacts (Wegiera et al., 2011; El-Shora et al., 2022). Regarding allelopathy, some species of *Rumex* were tested with seed germination, plant growth, weed control, insects, and herbivore activities. Many secondary metabolites have been in the species of this genus that researchers reported allelopathic conclusions on other plants and organisms (Costan et al., 2023). Despite *Rumex*, some other individuals of the Polygonaceae family possess allelopathic properties. *Polygonum* and *Fagopyrum* (buckwheat) have been well investigated for their repressive influences on the germination and growth of adjacent plants. For example, *Fagopyrum* suppresses weeds like *Amaranthus retroflexus* via secretion of phenolic acids and flavonoids into the root zone (Golisz et al., 2007). Comparably, *Polygonum hydropiper* has strong allelopathic consequences on crop plants and weeds because it contains allelopathic bioactive compounds (Wang et al., 2009). These signify that the Polygonaceae family is characterized by prevalent allelopathic activity, which helps in both ecological competitiveness and application in sustainable agriculture. Therefore, in the current review, the results of the investigation on the allelopathic effect of *Rumex* spp. will be presented and discussed that have been studied.

**Allelochemicals in *Rumex* spp.**

Allelochemicals are compounds naturally produced by plants and affect the growth and germination of neighboring plants after release. Secondary metabolites are at the top of the chemicals that allelopathic effects are attributed to, including phenols, alkaloids, terpenes, non-proteinaceous amino acids, glycosides, and sugars (Lin et al., 2007). These phytochemicals can be detected in the roots, stems, leaves, flowers, fruits, and seeds of allelopathic plants (Zeng, 2008). Allelochemical interactions comprise the production and elimination of reactive oxygen species (ROS) and the cell redox state. The recipient plant may form ROS after being subjected to allelochemicals in the regions of connection and influence the antioxidant activity of enzymes (Ding et al. 2007). Furthermore, Shearer et al. (2012) elaborated on the mechanism of allelochemical action on the target plant, and they deemed that signal transduction changes occurred along with an imbalance between ROS formation and antioxidant competences. Several species of *Rumex* were assessed for phytochemical profiles, including allelochemicals. In this context, Fairbairn and El-Muhtadi (1972) evaluated 19 species of *Rumex*; emodin, chrysophanol, and physcion were found in all of them; however, some of them were aloe-emodin and others were rhein or rhein-like compounds. *Rumex crispus* was examined for its allelopathic effect, and chromatography analysis established the occurrence of three phenolics, which inhibited seed germination of sorghum and radish (Einhellig and Rasmussen, 1973). Choi et al. (2004) isolated chrysophanol, parietin, and nepodin from *Rumex crispus*. In *Rumex japonicus*, Elzaawely et al. (2005) found protocatechuic, p-hydroxybenzoic, syringic, ferulic acids, vanillin, pyrocatechin, and caffeic acid in the analyzed root and shoot by high-performance liquid chromatography (HPLC) and gas chromatography–mass spectrometry (GC-MS). Besides, GC-MS analysis confirmed the occurrence of alkaloids, flavonoids, phenols, and glycosides in *Rumex vesicarius* (Alazzam et al., 2021). *Rumex dentatus* was investigated by researchers, and a vast number of bioactive chemicals were detected. The rain leaches of *Rumex dentatus* ssp. *klotzschianus* were inspected to find allelochemicals, and caffeic, p-coumaric, p-hydroxybenzoic, chlorogenic, syringic, ferulic, vanillic, and ocoumaric acids were detected (Hussain et al., 1997). Other researchers observed many different compounds in *Rumex dentatus*, such as chrysophanol, physcion, emodin, endocrocin, and chromones (Zhu et al., 2006); alkaloids, saponins, anthraquinones, and tannins (Fatima et al., 2009); and 1-dodecanamine, N,N-dimethyl-, 1-tetradecanamine, N,N-dimethyl, 2,6,10-trimethyl, 14-ethylen-14-pentadecne, n-hexadecanoic acid, 3-(N-Benzyl-N-methyl-amino)-1,2-propanediol, and cis-vaccenic acid (El-Shora et al., 2022). Moreover, Costan et al. (2023) analyzed leaves and roots of three *Rumex* species (*R. obtusifolius, R. crispus, and R. conglomeratus*) for allelopathic chemicals and found oxalates, phenols, and tannins that are essentially known for their allelopathic effect. Two allelopathic compounds with inhibitory effects were isolated from *Rumex maritimus* by Islam et al. (2017), which are 5,7-dihydroxyphthalide and altechromone A.

**Allelopathic effects of *Rumex* spp. are extract concentration dependent**

At the beginning of coining the term allelopathy by Molisch (1937), he was aware that any allelopathic substance that has inhibitory consequences at a certain concentration may have stimulatory at a lower one. For that reason, Molisch used the term allelopathy to cover both inhibitory and stimulatory effects (Willis, 2007). Subsequently, the term hormesis was ascribed to this dual phenomenon (Duke et al., 2006). The physiological mechanism behind hormesis may be due to at low doses, the allelochemicals induce production growth promoter agents, including oxygen-free radicals to boost defense mechanisms, increasing plant hormones from growth promoter classes, and induction of growth-improving related pathways. In contrast, at high doses, the allelochemicals reverse the latter mechanisms and inhibit the target plant growth (Parsons, 2003; Kovalchuk et al., 2003). For the allelopathic effect of *Rumex* spp. according to the extract concentration, Pilipapivičius et al. (2012) observed an increase in germination percentage of spring barley at low concentrations of *Rumex crispus* (0.01 g/Petri dish), while at high concentrations (0.25 to 0.5 g/Petri dish) germination was inhibited. In their study, Khalid et al. (2021) used fresh and dry leaf extracts of *R. dentatus* at various concentrations (5, 10, and 15 g/100 mL of water) to *Brassica campestris* and concluded that the inhibitory effect of *R. dentatus* on growth and germination was increased as concentration increased, particularly from dry leaves. Additionally, five concentrations of *R. dentatus* (0%, 2%, 4%, 6%, and 8% w/v) were applied to seeds, and it was revealed that germination inhibition was very pronounced as concentration elevated (Mukhtar et al., 2012). In another study, shoots and roots of *R. dentatus* were collected at vegetative, flowering, and fruiting stages, and extracts were prepared from them at (0%, 20%, 40%, and 60%), then exposed to seeds of wheat *in vitro*. Shoot extracts at 20% from vegetative and flowering stages showed a slight elevation in seed germination but declined at higher concentrations. Root extracts decreased germination at all concentrations (El-Beheiry et al., 2019). Allelopathic activity of different extract concentrations of *Rumex japonicus* was examined on lettuce and barnyard grass seeds; the obtained results disclosed obvious inhibitory allelopathy of *R. japonicus* extracts at 500 ppm (Elzaawely et al., 2005). In an additional study, Islam et al. (2017) investigated varying extract concentrations (0.01, 0.1, 0.03, and 0.3 g DW$^{-1}$ extract equivalent/mL) of *Rumex maritimus* and their detected allelochemicals (5,7-dihydroxyphthalide and altechromone A) for their allelopathy phenomenon, and they concluded that the effect of the extracts and their allelochemicals was concentration-dependent; extreme inhibition was recorded at (0.3 g $DW^{-1}$ extract equivalent/mL) in the shoot and root growth of cress, rapeseed, barnyard grass, and foxtail fescue. In a somewhat different study, Islam et al. (2024) used *R. maritimus* residues in the field of rice at (0, 1.0, 2.0, 3.0 t $ha^{-1}$). *R. maritimus* residues at 3.0 t $ha^{-1}$ significantly reduced weed growth and had a promising impact to enhance rice yield. Furthermore, an allelopathic bioassay of *Rumex vesicarius* was conducted on wheat seeds at (0.5%, 1%, and 2%); the obtained results clarified that an aqueous extract concentration of 2% reduced wheat germination to 13.3% (Alazzam et al., 2021). Increasing ROS and enzyme activities in the target plant were related to the extract concentration of the allelopathic plant. El-Shora et al. (2022) demonstrated high ROS (hydrogen peroxide, superoxide radicals, and hydroxyl radicals) and activities of enzymes (glucose-6-phosphate dehydrogenase, 6-phosphoglucnate dehydrogenase, nitrate reductase, and glutamine synthetase) in leaves of *Portulaca* plants by increasing the concentration of *R. dentatus* leaf extract.

**Allelopathic capacity of *Rumex* spp. depending on the plant part**

Allelochemicals can be found in the roots, stems, bark, leaves, fruits, and seeds of various plants. These plant organs have different capacities to produce and store the allelochemicals (Hussain et al., 2011; Ahmad et al., 2020; Ibrahim et al., 2025). Perhaps allelochemicals are leached into the soil from leaves, stems, and bark by rain, or the chemicals are released into the soil after the decomposition of the fallen leaves; on the other hand, roots are well-known for releasing allelochemicals into the soil and targeting the recipient plants (Inderjit and Duke 2003). This variation in allelopathic potential among plant organs of the same plant presumably belongs to the differences in pathways that occur in these organs or the ability of the organs to store the allelochemicals (Grodzinskii, 2016). Regarding roots and leaves of three *Rumex* species (*R. obtusifolius, R. crispus*, and *R. conglomeratus*), they were assessed for allelochemicals, oxalates, phenols, and tannins; among them, roots had more tannins than leaves (Costan et al., 2023). In other species of *Rumex* using extracts from different plant parts, Gam et al. (2024) measured half-maximal inhibitory concentration in g $L^{-1}$ ($IC_{50}$) in shoot and root extracts of *Rumex acetosella* for germination of *Trifolium repens* seeds. The value was higher in the root extract (1.72 g $L^{-1}$) than in the shoot extract (1.31 g $L^{-1}$); thus, the shoot extract was more inhibitory than the root extract. Besides, aerial parts and roots of *R. japonicus* were studied for allelopathic effect using lettuce and barnyard grass seeds as a bioassay. It was evident from the results that ferulic acids existed in the root but not in aerial parts, pyrocatechin was higher in the aerial parts than in the roots, and growth inhibition by the root extract was greater than by aerial parts (Elzaawely et al., 2005). Various concentrations of *R. crispus* were prepared from roots, shoots, and seeds and tested with seeds of spring barley for allelopathic effect. The maximum germination suppression was found due to shoot extracts rather than root extracts, but seed extract had a stimulatory influence on seed germination (Pilipapivičius et al., 2012). Hussain et al. (1997) reported in their results about the allelopathic effect of *R. dentatus* ssp. *Klotzschianus* on wheat and mustard that fresh and dried root extracts of *R. dentatus* ssp. *Klotzschianus* were more inhibitory than fresh and dried shoot extracts in both boiled and unboiled water (5 g/100 mL). Additionally, the allelopathic effect of plant parts may be related to the growth stage of the allelopathic plant. Root and shoot allelopathic impacts of *R. dentatus* at vegetative, flowering, and fruiting stages were investigated on wheat. At low concentrations, shoot extracts at vegetative and flowering stages had a stimulatory result on germination, whereas shoot extracts at high concentrations and root extracts at all concentrations that were collected at the three growth stages had inhibitory consequences on wheat germination and growth (El-Beheiry et al., 2019).

**Response of different plant species to allelopathic effects of *Rumex* spp.**

An allelopathic plant may have an allelopathic effect on a wide range of plants. However, a species-specific effect of an allelopathic plant has been recorded. Different plant species may respond differently to the same allelopathic plant at the same extract concentrations (Bieberich et al., 2018). The reason why plant species and varieties respond differently to the same plant allelopathic effect is likely due to plants having various genetic makeups and physiological traits, including permeabilities of the membranes and metabolic pathways that interfere with the sensitivity of the owner plant to allelochemicals (Das and Kato-Noguchi, 2018). In this connection, *Amaranthus retroflexus* L. (pigweed), sorghum, and corn were exposed to the allelopathic effect of *R. crispus* as a leaf extract, and growth inhibition in the three species was observed at all concentrations of *R. crispus*, but corn was not affected by the lowest (most diluted) concentration (Einhellig and Rasmussen, 1973). Testing allelopathic outcomes of *R. dentatus* with seeds of wheat and mustard revealed that dried shoot extracts of *R. dentatus* in boiled water and fresh shoot extract in unboiled water reduced germination and plumule growth in mustard but not in wheat (Hussain et al., 1997). In their study on the allelopathic effect of *R. japonicus* on lettuce and barnyard grass, Elzaawely et al. (2005) showed that the shoot and root of barnyard grass were less inhibited by *R. japonicus* extracts than lettuce. Moreover, Zaller (2006) categorized the leguminous and non-leguminous herb and grass species as affected and non-affected by the allelopathic effect of *R. obtusifolius*, and he declared that the tested species were species-specific susceptible. Two species of *Rumex* (*R. crispus* and *R. obtusifolius*) were assessed for their allelopathic impacts on *Medicago sativa*, *Trifolium pratense*, *Trifolium repens*, and *Lotus corniculatus*. Germinations of *Medicago sativa*, *Trifolium pratense*, and *Lotus corniculatus* were severely reduced by both *Rumex* species, but *Trifolium repens* was moderately affected (Camen et al., 2017). Seeds of cress, rapeseed, barnyard grass, and foxtail fescue were subjected to *R. maritimus* extracts; growth of cress, rapeseed, and foxtail fescue was completely (100%) inhibited by extract concentration 0.3 g $DW^{-1}$ of the *R. maritimus*. The same concentration reduced the growth of barnyard grass but to a lesser extent (Islam et al., 2017). Additionally, grain per panicle, grain yield, and biological yield of two varieties of rice (BRRI dhan58 and BRRI dhan74) differently responded to the same ratio (3.0 t $ha^{-1}$) of *R. maritimus* residues. The results explained that the highest values of the three measurements were observed in BRRI dhan74 rather than in BRRI dhan58 (Islam et al., 2024).

**Allelopathic effects of *Rumex* spp. on seed germination and growth**

Plant allelopathic effects can be either deleterious or favorable on seed germination and growth of the recipient plant. These are generally evaluated by several physiological mechanisms that cause germination and growth inhibition/stimulation (Mutlu and Atici, 2009; Tahir et al., 2022). Allelopathic plants may interfere with the storage and mobilization of the metabolites, respiration, photosynthesis, osmosis, etc., in the recipient plant. With the *in vitro* allelopathic effect of *Rumex crispus*, both stimulatory and inhibitory effects were observed on the seed germination of spring barley depending on concentration (Pilipapivičius et al., 2012). In another species of *Rumex*, Bhatia et al. (1982) reported that leaf extract of *R. dentatus* stimulated seedling length of wheat but declined dry weight. In the same *Rumex* species, Hussain et al. (1997) found *in vitro* that wheat and mustard germination was reduced by root extract of *R. dentatus* ssp. *Klotzschianus* that was extracted in hot water. Radicle growth of wheat and mustard was so sensitive to the extracts of the root and shoot of this *Rumex* species, whether they were dried or fresh and extracted in cold water or hot water. Whereas fresh and dried root extracts in both hot and cold water were inhibitory to plumule growth in wheat and mustard, and shoot extract was just an inhibitor for both species to plumule growth in hot water. Furthermore, seed germination inhibition in lettuce and barnyard grass was detected *in vitro* due to extracts from aerial parts of *R. japonicus* at all concentrations. In contrast, inhibition in root and shoot growth of lettuce and barnyard grass required the highest extract concentration of *R. japonicus*. Aerial part extract inhibited shoot growth at 500 ppm, and root growth was inhibited by root extract at 500 ppm (Elzaawely et al., 2005). In addition, fresh leaf extract of *R. obtusifolius* (1000 g $L^{-1}$) was investigated for allelopathic effect *in vitro* and in field experiments with some herbs and grasses. Seed germination of all grasses and some herbs was inhibited by *R. obtusifolius* extract *in vitro*; however, the extract spray on the field was not effective to inhibit germination in the recipient species. Also, germination promotion was not observed in the tested species by *R. obtusifolius* extract (Zaller, 2006). In an additional similar experiment, the fresh leaf of *R. crispus* extract was applied to *Amaranthus retroflexus*, sorghum, and corn *in vitro* and in the field. Regarding the *in vitro* experiment, growth reduction in the three species was detected, and also growth reduction was observed in some species near *R. crispus* compared to the ones away from this *Remex* species in the field (Einhellig and Rasmussen, 1973). Moreover, suppression of seed germination, radical and plumule lengths were recorded in *T. aestivum in vitro* by leaf powder of *R. dentatus* (Anwar et al., 2013). In an *in vitro* experiment, the extract of *R. maritimus* was applied to seeds of cress, rapeseed, barnyard grass, and foxtail fescue. *R. maritimus* extract inhibited the growth of the seedlings of the recipient plants in both root and shoot (Islam et al., 2017). Germination percentages of alfalfa, red clover, and white clover were declined under extracts of *R. crispus* and *R. obtusifolius* leaf extracts *in vitro* (Camen et al., 2017). A reduction in seed germination of wheat was found because of the application of *R. vesicarius* as methanolic and aqueous extracts. Besides, high concentrations of both extracts reduced radicle and plumule growth (Alazzam et al., 2021). Extract from leaves of *R. dentatus* adversely affected germination percentage, radicle, and plumule growth of *Brassica campestris*, and plumule was the most affected part (Khalid et al., 2021). In another *in vitro* study, Heděnec et al. (2014) exposed seeds of wheat and mustard in a petri dish containing sand and soil to extract of *Rumex tianschanicus* x *Rumex patientia* hybrid. The results proved that the extract of this hybrid decreased seed germination of wheat and mustard in both substrates.

**Possibility of control weeds by *Rumex* spp. allelopathy**

Weeds are a great problem during the production of different agricultural crops. For weed management, herbicides are still one of the options that are widely used, but herbicides are hazardous to humans, animals, and the environment. Therefore, efforts have been exerted to find effective alternatives to chemical herbicides, including allelochemicals from allelopathic plants. *Rumex* spp. are among the species that have been frequently studied for allelopathic effects on weeds. Camen et al. (2017) used extracts from *R. crispus* and *R. obtusifolius* for seeds of grassland legumes, and they recorded a reduction in germination of those legumes with different extents depending on species. Besides, El-Shora et al. (2022) found an increase in total phenolics and total flavonoids along with a reduction in metabolic activities and macromolecule contents in leaves of *Potulaca oleracea* because of the application of *R. dentatus* leaf extract. They referred to these reductions as providing a stressful condition for the growth of *P. oleracea*, and this will be a promising base for using *Rumex* leaf extract as a bioherbicide. Furthermore, the results of using *R. maritimus* residues in the field of *boro* rice clarified that the yield of *boro* rice was improved due to inhibition of weed growth by *R. maritimus* residues. As a result, *R. maritimus* residues can be used as an option for weed control in sustainable crop production (Islam et al., 2024). Moreover, Gam et al. (2024) detected that *R. acetosella* shoot extract impeded the growth of white clover, and they reported that this will be an alternative strategy for weed management. Despite all the findings, the big problem with using allelopathic plant extracts as herbicides in the majority of the cases is non-selectivity; both cultivated crops and weeds are affected by the allelopathic plant extract. Islam et al. (2017) declared that inhibition in seedling growth of cress and rapeseed, which are two crops, along with inhibition in seedling growth of the two weeds barnyard grass and foxtail fescue. Hussain et al. (1997) found inhibition in germination and seedling growth of wheat along with mustard weed as a result of the allelopathic effect of *R. dentatus* ssp. *Klotzschianus*. For that reason, application extract of allelopathic plants before crop cultivation may be promising for weed management.

**Insecticidal activity of *Rumex* spp. allelopathy**

Using synthetic insecticides negatively influences humans, animals, and the environment and also stimulates insect resistance to insecticides. These force the researchers to find out natural alternatives. In such a setting, an aqueous extract of *Rumex nepaiensis* at 5% was applied to reduce diamondback moth, cabbage white butterfly, and cabbage aphid. The results indicated a reduction in feeding and settlement of diamondback moth and cabbage white butterfly (Mehta et al., 2005). Additionally, El-Badawy et al. (2021) applied leaf methanolic extract of *R. dentatus* at 3%, 5%, and 7% against saw-toothed grain beetle, rice weevil, and bean weevil stored grain pests. They recorded the maximum antifeedant effect against saw-toothed grain beetle and strong adult emergence inhibition of the three pests due to methanolic extract of *R. dentatus* at the three concentrations. In addition, to reduce insect resistance toward *Bacillus thuringiensis* delta-endotoxins, Mhalla et al. (2018) combined *Bacillus thuringiensis* BLB250 with *Rumex tingitanus* hexane extract and used it on *Spodoptera littoralis* larvae, and they observed a significant synergistic effect against *S. littoralis* larvae in the combination of BLB250 delta-endotoxins and hexane extract of *R. tingitanus*. Also, they detected β and γ-sitosterol, campesterol, and β-amyrin with insecticidal activities in GC-MS analysis of *R. tingitanus* hexane extract. In another study, stem and leaf crude methanolic extracts and n-Hexane, chloroform, ethyl acetate, and methanol fractions of *Rumex nervosus* were tested for insecticidal potential against *Rhyzopertha dominica* and *Tribolium castaneum*. Low mortality was observed just against *R. dominica* owing to leaf crude methanolic extract and chloroform and ethyl acetate fractions; other treatments were inactive (Khan et al., 2018). Hussain et al. (2010) revealed the insecticidal activity of *R. dentatus* and *Rumex nepalensis* but not *Rumex hastatus,* as they used crude extracts against *Sitophilus* *oryzae*. In a somewhat different study, Salama et al. (2022) reported that methanolic shoot extract of *R. vesicarius* had a great capacity to control disease vector insects.

**Antibacterial and antifungal activities of *Rumex* spp. allelopathy against plant diseases**

Plant diseases are in most cases controlled by synthetic chemicals. Nowadays, many serious problems limit using those chemicals to control plant diseases. Chemicals result in adverse and fatal effects on non-target organisms, developing resistance in pathogens, and they are expensive. Accordingly, a natural, affordable, and environmentally friendly substitute needs to be used (Da et al., 2019). In their attempt, Choi et al. (2004) showed that the methanol extract of the *R. acetocella* root declined powdery mildew development in barley *in vivo*. Besides, Kim et al. (2004) proved methanolic root extract of *R. crispus* as an effective solution to control barley powdery mildew at 11 g $L^{-1}$ Tween 20 *in vivo*. Also, at 300 g $L^{-1}$ Tween 20, *R. crispus* root extract was influential against *Sphaerotheca fuliginea* on cucumber like a fenarimol (30 mg $L^{-1}$) fungicide and more than polyoxin B (100 and 33 mg $L^{-1}$) fungicide. On the other hand, the capacity of the antibacterial and antifungal activities of *Rumex* spp. is varying depending on the *Rumex* species and species of the fungus and bacterium. Hussain et al. (2010) demonstrated that *R. hastatus* leaf extract and *R. nepalensis* root extract had low antibacterial activity against *Pseudomonas aeruginosa* plant bacterium, but *R. dentatus* leaf extract had moderate activity against the same bacterium. The same researchers also showed that *R. hastatus* leaf extract had intermediate inhibitory activity against *Helminthosporium maydis* and *Aspergillus niger*, and low activity against *Fusarium solani*, *Aspergillus flavus*, and *Alternaria solani* fungi. Leaf extract of *R. dentatus* acted moderately against *F. solani* and poorly against *A. flavus*, *A. niger*, *A. solani*, and *H. maydis* fungi. *R. nepalensis* root extract had the maximum activity against *A. niger*, intermediate against *A. flavus* and *A. solani*, but was inactive against *F. solani* and *H. maydis*. In the framework of studies that have been done on the allelopathic effect of *Rumex* spp. on the growth of plant pathogens, Heděnec et al. (2014) exhibited less growth of fungal pathogens in soil as a result of treatment of those fungi on agar with the leachate of the sorrel hybrid *R. tianschanicus* x *R. patientia*. In an *in vitro* study, Alotibi et al. (2020) applied shoot extract of *R. vesicarius* against *Fusarium*, *Helminthosporium*, *Alternaria*, and *Rhizoctonia* species, and the results revealed antifungal activity of the extract and suppressed sporogenesis of *Alternaria* and *Fusarium* and changed the morphology of the hyphal shape and surface of the fungi. Furthermore, Pokhrel and Choden (2022) concluded that the aerial part extract of *R. nepalensis* was so efficient against *A. solani* on potato than the root extract of the same species. They reported that the highest concentration was the best and suggested aerial part extract of *R. nepalensis* as a bio-fungicide for the control of *A. solani* in potato. Rashid et al. (2024) used *Rumex tuberosus* aqueous extract to seeds of tomato and indicated that the extract at 15% was active against seedborne pathogens.

## Conclusion

In the current review, the allelopathic effect of about 14 *Rumex* species was reviewed that have been studied by researchers on various aspects of plant and other plant-related problems. In this context, *Rumex* spp. were rich in allelochemicals, and these allelochemical contents were different according to the species of the *Rumex*. The allelopathic effect of the studied *Rumex* species depended on extract concentration. Both stimulatory and inhibitory effects of *Rumex* allelopathy were observed on the recipient plant depending on concentration and the species of the *Rumex*. In the species that had the dual effects, at low concentrations they were stimulators, but at high concentrations they were inhibitors. However, in some of the *Rumex* spp., they were inhibitors at all studied concentrations. Besides, the allelopathic effect of the *Rumex* spp. was highly dependent on the plant part. Some species have strong inhibitory allelopathy in roots but others in shoots. Seeds contain no inhibitory effect but stimulatory effects. Besides, the allelopathic effects of the plant parts were variable depending on the stage of the growth. Furthermore, the allelopathic effect of the *Rumex* spp. was not the same on different plant species or varieties. Additionally, both positive and negative effects of *Rumex* spp. allelopathy were observed on seed germination and growth of the recipient plants. Promising results have been obtained with *Rumex* spp. allelopathy to weed management, whereas both crops and weeds were influenced by *Rumex* allelopathy. Hence, it is better to use *Rumex* allelopathy to control weeds before cultivation. Also, the allelopathic effect of *Rumex* spp. against insects and plant pathogens was promising. Inhibitions in insects and plant pathogens were obtained with the allelopathic effects of *Rumex* spp. In spite of all the findings, there is little information on the field trials of *Rumex* spp. allelopathy; most of the studies were about *in vitro* settings, particularly for control weeds. A large number of *Rumex* species have remained to study their allelopathic impact.